\documentclass[twocolumn,showpacs,preprintnumbers,amsmath,amssymb,latexsym,prl]{revtex4-1}
\usepackage{latexsym,amssymb,amsfonts,amsmath}
\usepackage{color}
\usepackage[dvips]{graphicx}

\newcommand{\TK}[1]{\textcolor{blue}{#1}}

\usepackage{graphicx}% Include figure files
\usepackage{dcolumn}% Align table columns on decimal point
\usepackage{bm}% bold math
\usepackage{ulem}

\begin{document}

%\preprint{APS/123-QED}

\title{Universality and the QCD Anderson Transition}\thanks{Supported
    by the Hungarian 
    Academy of Sciences under ``Lend\"ulet'' grant
    No. LP2011-011. TGK and FP acknowledge partial support by the EU
    Grant (FP7/2007 -2013)/ERC No. 208740. We also thank the
    Budapest-Wuppertal group for allowing us to use their code to
    generate the gauge configurations.}

\author{Matteo Giordano}\author{Tam\'as G.\ Kov\'acs}\author{Ferenc Pittler }

\affiliation{%
Institute for Nuclear Research of the Hungarian Academy of Sciences, \\
H-4026 Debrecen, Bem t\'er 18/c, Hungary %\textbackslash\textbackslash
}%

\date{\today}% It is always \today, today,

\begin{abstract}

We study the Anderson-type transition previously found in the spectrum
of the QCD quark Dirac operator in the high temperature, quark-gluon
plasma phase. Using finite size scaling for the unfolded level spacing
distribution, we show that in the thermodynamic limit there is a
genuine mobility edge, where the spectral statistics changes from
Poisson to Wigner-Dyson statistics in a non-analytic way. We determine
the correlation length critical exponent, $\nu$, and find that it is
compatible with that of the unitary Anderson model. 

\end{abstract}

\pacs{12.38.Gc,72.15Rn,12.38.Mh,11.15.Ha}% PACS, the Physics and Astronomy
\maketitle

The idea of Anderson localization is more than half a century
old~\cite{Anderson58}. Anderson localization consists in the spatial
localization of the states of a system due to quantum interference
effects, caused by the presence of disorder. Its simplest realization
is provided by the Anderson tight-binding model that aims at
describing electronic states in a ``dirty'' conductor, by mimicking
the effect of impurities through a random on-site potential.  In three
dimensions, as soon as the random potential is switched on, localized
states appear at the band edge. However, states remain extended around
the band center, beyond a critical energy called the ``mobility
edge''. Increasing the amount of disorder, i.e., increasing the width
of the distribution of the random potential, the mobility edge moves
towards the band center, and above a certain critical disorder all the
states become localized (see Refs.~\cite{Lee:1985zzc,Evers:2008zz}). 

Originally proposed to explain the loss of zero temperature conductance as a
result of disorder, localization was later found in a much wider range of
physical systems. Anderson transitions have been demonstrated with
electromagnetic and sound waves as well as cold atoms (see
Ref.~\cite{KMOS} and references therein) and recently in strongly
interacting matter in its high temperature quark-gluon plasma
phase~\cite{Kovacs:2012zq}. The last item of the list is rather
peculiar since in that case localization occurs on a vastly different
length and energy scale from all previously known cases, namely on
subnuclear rather than atomic scales.

In the microscopic description of strongly interacting matter provided by
quantum chromodynamics (QCD), a central role is played by the Dirac
operator. \TK{Its} spectrum encodes important properties of quarks and
hadrons. At low temperature, the lowest lying quark eigenmodes of the Dirac
operator have long been known to be extended, and the corresponding spectrum
to obey Wigner-Dyson statistics as predicted by random matrix theory
(RMT)~\cite{Verbaarschot:2000dy}.  This has been successfully exploited to
study the low-energy properties of QCD~\cite{Verbaarschot:2000dy}. In
contrast, in the high-temperature quark-gluon plasma phase no similar
description of the low-lying quark modes was available until recently. It was
first suggested by Garc\'ia-Garc\'ia and Osborn that the transition from the
hadronic to the quark-gluon plasma phase might be an Anderson-type
transition~\cite{GarciaGarcia:2006gr}. Using lattice QCD they qualitatively
demonstrated that heating the system through the critical temperature makes
the quark states more localized. However, at that time a detailed verification
of an Anderson transition in QCD was not possible.

More recently, using lattice simulations at a fixed temperature well above the
crossover temperature, $T_c$, we explicitly verified the existence of an
Anderson-type transition in the spectrum of the quark Dirac operator. We found
that while the lowest part of the spectrum consists of localized states that
obey Poisson statistics, higher up in the spectrum the states become
delocalized and the level spacings are described by Wigner-Dyson random matrix
statistics~\cite{Kovacs:2012zq}. The scaling of the mobility edge, separating
localized and delocalized states, indicates that it survives in the continuum
limit and it steeply increases with the temperature.  Thus the temperature in
QCD plays a role similar to the disorder strength in the Anderson model. We
also demonstrated that in larger volumes the transition becomes sharper,
suggesting that it is a real phase transition.

In this letter we present a finite size scaling study of the transition and
show explicitly that it is a genuine second order phase transition. We also
compute the correlation length critical exponent, $\nu$, and find that it is
compatible with that of the three-dimensional Anderson model in the unitary
class, the class to which quarks in the fundamental representation of the
SU(3) gauge group are also expected to belong. Our results suggest that the
universality of the Anderson transition might be much more general than
previously thought. So far, universality in the Anderson model had been
checked for different distributions of the diagonal disorder (see
e.g.~\cite{nu_latest}) and for uncorrelated off-diagonal
disorder~\cite{offd_disorder}. The model we consider here, lattice
QCD, is very different from all previously considered cases. Here the
disorder appears 
through the gauge fields in the hopping terms, while the on-site terms are
identically zero. Moreover, the random fluctuations of the disorder at
different locations are not independent. However, since the theory has a mass
gap, correlations among them decay exponentially with the distance. It is also
remarkable that in QCD the transition is not driven by the disorder strength
but by the temperature. In lattice QCD the temperature is set by the
extension of the system in Euclidean time, as $T=1/L_t$, where $L_t$ is the
temporal size. As the system is heated and $L_t$ becomes smaller, the lowest
lying quark modes are squeezed not only in the temporal but also in the
spatial directions. This results in the localization of the lowest quark
modes, up to the mobility edge, which in turn is pushed to higher values as
the system is further heated. For a possible physical explanation of this
mechanism in terms of the antiperiodic temporal quark boundary condition and
fluctuations of the Polyakov loop, see Ref.~\cite{Bruckmann:2011cc}.

\begin{figure}[t]
  \centering
  \includegraphics[width=0.43\textwidth]{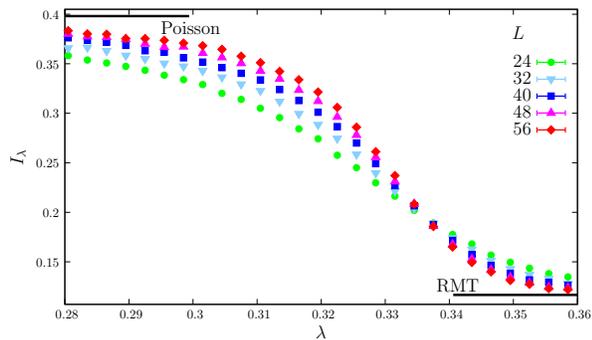}
  \caption{Integrated ULSD as a function of $\lambda$ for several
    lattice sizes. Here
    $\Delta\lambda\cdot 10^3=3$.}
  \label{fig:0}
\end{figure}

More precisely, the model we consider here is lattice QCD at finite
temperature, with 2+1 flavors of staggered quarks with the quark masses tuned
to the physical $u,d$ and $s$ quark masses. Here QCD is discretized on a 3+1
dimensional hypercubic lattice with three spatial and one Euclidean temporal
dimension with the temporal size setting the physical temperature of the
system. For details of the particular action and parameters we use, see
Refs.~\cite{Aoki:2005vt} and \cite{Kovacs:2012zq}. The staggered Dirac
operator is a simple lattice discretization of the continuum Dirac operator
containing covariant derivatives with the SU(3) color gauge field. Technically
the staggered Dirac operator is thus a large sparse matrix with all zeros in
the diagonal (on-site terms) and non-zero elements only in the nearest
neighbor hopping terms. Being a discretized covariant derivative, each hopping
term depends on the SU(3) group valued gauge field attached to the given link
(parallel transporter). The gauge links, in turn, are random variables
generated with the full QCD path integral measure (see
e.g.\ Ref.~\cite{Montvay:1994cy}).

The spectrum of the staggered Dirac operator on a finite lattice is a discrete
set of pairs of purely imaginary eigenvalues $\pm i\lambda_n$. Here and in the
following, $\lambda$ denotes the eigenvalues in lattice units. For our
purposes it is enough to restrict to the positive part of the spectrum
$\lambda_n\ge 0$. At temperatures above $T_c$, the lowest-lying eigenmodes of
the Dirac operator are localized on the scale of the inverse temperature,
while higher up in the spectrum they are delocalized. For the present study,
we use a fixed temperature of $T\simeq 2.6~T_c$, corresponding to temporal
extension $L_t=4$ in lattice units, and lattice spacing $a=0.125~{\rm
  fm}$. The temporal size of the system is thus fixed and we vary only its
size in the three spatial dimensions, using linear extensions of
$L=24,28,32,36,40,44,48,56$ (in lattice units).  Our results are based on a
rather high statistics for present day lattice QCD standards, consisting of
40k independent configurations on the smallest lattice; on the larger
lattices the number of configurations was scaled down with the volume
to have the same eigenvalue statistics.

% \begin{figure}[t]
%   \centering
%   \includegraphics[width=0.43\textwidth]{fig1.eps}
%   \caption{Integrated ULSD as a function of $\lambda$ for several
%     lattice sizes. Here
%     $\Delta\lambda\cdot 10^3=3$.}
%   \label{fig:0}
% \end{figure}

\begin{figure}[t]
  \centering
  \includegraphics[width=0.42\textwidth]{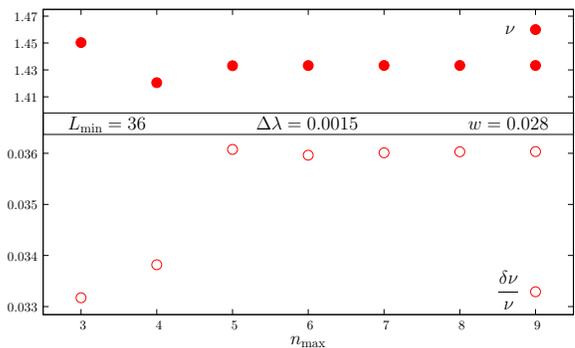}
  \caption{The fitted value of $\nu$ and the corresponding relative error
    versus the number of terms $n_{\rm max}$, in the case of $L_{\rm min}=36$,
    $\Delta\lambda\cdot 10^3=1.5$ and fitting range width $w\cdot 10^2=2.8$.}
  \label{fig:bay_conv}
\end{figure}

A convenient way to investigate the transition from localized to delocalized
modes is to study the local statistical properties of the spectrum, which are
expected to show a change from Poisson to RMT behavior across an Anderson
transition. In this respect, a possible quantity to consider is the so-called
unfolded level spacing distribution (ULSD), which is known analytically for
both kinds of statistics. Unfolding is a local rescaling of the eigenvalues to
have unit spectral density throughout the spectrum. The ULSD gives the
probability distribution of the difference between two consecutive eigenvalues
of the Dirac operator normalized by the local average level spacing. In the
thermodynamic limit, the critical point (mobility edge) in the spectrum,
$\lambda_c=\lambda_c(T)$, separating localized and delocalized modes, is
identified as the point where the local ULSD $P_\lambda(s)$ switches between
Poisson and Wigner-Dyson statistics.

Any quantity extracted from the local ULSD, having different values for
Poisson and RMT statistics, can be used to detect the transition, and to study
the corresponding critical behavior, along the lines of
Refs.~\cite{HS,SSSLS,SP}. Denoting by $Q(\lambda,L)$ such a quantity, computed
on a lattice of linear size $L$, one expects for its thermodynamic limit
$Q(\lambda)=\lim_{L\to\infty} Q(\lambda,L)$ the following behavior:
$Q(\lambda)= Q_{\rm Poisson}$ for $\lambda<\lambda_c$;
$Q(\lambda_c)=Q_c$, and $Q(\lambda)=Q_{\rm RMT}$ for $\lambda>\lambda_c$.
In a second-order phase transition such as the Anderson transition, the
characteristic length of the system, $\xi_\infty$, diverges at the critical
point like $\xi_\infty(\lambda)\sim |\lambda-\lambda_c|^{-\nu}$. Close
to $\lambda_c$ and in large enough volumes, so that corrections to
one-parameter scaling can be ignored, finite size scaling suggests that the
dependence of $Q$ on $L$ is of the form $Q(\lambda,L) =
f(L/\xi_\infty(\lambda))$.  As $Q(\lambda,L)$ is analytic in $\lambda$ for any
finite $L$, we must have $ Q(\lambda,L)=F(L^{1/\nu}(\lambda-\lambda_c))$, with
$F$ analytic. This means that the data for different volumes, when plotted
against the scaling variable $L^{1/\nu}(\lambda-\lambda_c)$, should collapse
on a single scaling curve, $F$.

The parameters $\lambda_c$ and $\nu$ can be obtained by optimizing data
collapse for a set of volumes in the following way.  Expanding the scaling
function $F$ in powers of $\lambda-\lambda_c$,
\begin{equation}
  \label{eq:fss_expansion}
 Q(\lambda,L)=\sum_{n=0}^{\infty} F_{n}\,L^{n/\nu}(\lambda-\lambda_c)^n \,,
\end{equation}
one can truncate the series to an order $n_{\rm max}$, high enough to give a
good description of the scaling function in a range of width $w$ around
$\lambda_c$.  We can then fit the coefficients of the polynomial and the
parameters $\lambda_c$ and $\nu$ to the data on a set of volumes. The goodness
of the fits measures how precisely data collapse occurs. For the fit we used
the MINUIT library~\cite{James:1975dr}, and determined statistical errors by
means of a jackknife analysis.

For our purposes, the best choice for $Q$ turned out to be the integrated
ULSD, defined locally in the spectrum, $I_\lambda = \int_0^{s_0}
ds\,P_\lambda(s)$. Here $s_0\simeq 0.508$ was chosen in order to maximize the
difference between the values predicted by Poisson and RMT statistics, namely
$I_{\rm Poisson}\simeq 0.398$ and $I_{\rm RMT}\simeq 0.117$.  In practice,
$I_\lambda$ was computed by dividing the full spectrum in bins of width
$\Delta \lambda$, integrating the ULSD in each bin, and assigning the
resulting value to the average value of $\lambda$ in each bin. In
Fig.~\ref{fig:0} we show this quantity as a function of $\lambda$ for several
system sizes. % ranging from $24^3$ to $56^3$.

\begin{figure}[t]
  \centering
  \includegraphics[width=0.42\textwidth]{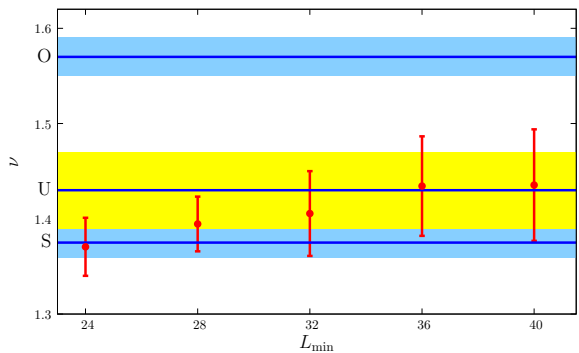}
  \caption{Dependence of the fitted value of $\nu$, averaged over
    $2.6\le w\cdot 10^2 \le 3$ and $1\le \Delta\lambda\cdot 10^3 \le 3$, on
    $L_{\rm min}$. The critical exponents
    measured for the orthogonal (O)~\cite{corrections},
    unitary (U)~\cite{nu_unitary} and symplectic (S)~\cite{nu_symp} Anderson
    models, and the corresponding error bands, are also shown for
    comparison.} 
  \label{fig:lmin_dep}
\end{figure}

\begin{table}[t]
  \centering
  \begin{tabular}{cccc}
    \hline    $L_{\rm min}$ & $\nu$ & $\lambda_c$ & $I_{\lambda=\lambda_c}$ \\
    \hline \hline
    24   &     1.371(30)  &      0.33637(13)  &0.19429(72) \\ 
    28   &     1.394(29)  &      0.33633(16)  &0.19451(97) \\ 
    32   &     1.405(44)  &      0.33626(18)  &0.1950(11)$\phantom{0}$  \\ 
    36   &     1.434(52)  &      0.33637(24)  &0.1943(16)$\phantom{0}$  \\ 
    40   &     1.435(59)  &      0.33604(35)  &0.1966(25)$\phantom{0}$  \\ 
    \hline
  \end{tabular}
  \caption{Fitted values of $\nu$, $\lambda_c$, and the integrated
    ULSD $I_{\lambda}$ at criticality, and corresponding
    errors as a function of $L_{\rm min}$.}
  \label{tab:1}
\end{table}

The quality of the fit reflects the goodness of the data collapse only if the
truncation of Eq.~\eqref{eq:fss_expansion} can provide a good description of
the scaling function in the required range. To check this we included more and
more terms in the series and monitored the stability of the results.  In order
to circumvent the numerical instability of polynomial fits of large order, we
resorted to the technique of constrained fits~\cite{Lepage:2001ym}. The basic
idea of constrained fits is to use the available information to constrain the
values of the fitting parameters. In our case, they are needed only to avoid
redundancy in the fitting parameters and the resulting instability of the
fits. We did not impose any constraint on $\lambda_c$, $\nu$ and $F_n$ for
$n\leq 3$. Our constraints on the higher order coefficients were also very
loose. In Fig.~\ref{fig:bay_conv} we plot the dependence of $\nu$ and its
uncertainty on the order of the truncation used for the fits. Both the value
and the error are absolutely stable for $n_{\rm max} \geq 5$, and in fact even
from $n_{\rm max} \geq 3$ changes are within the uncertainties. After
stabilization, the resulting errors include both statistical effects and
systematic effects due to truncation~\cite{Lepage:2001ym}.
In the following we use $n_{\rm max}=9$.

\begin{figure}[t]
  \centering
  \includegraphics[width=0.3\textwidth]{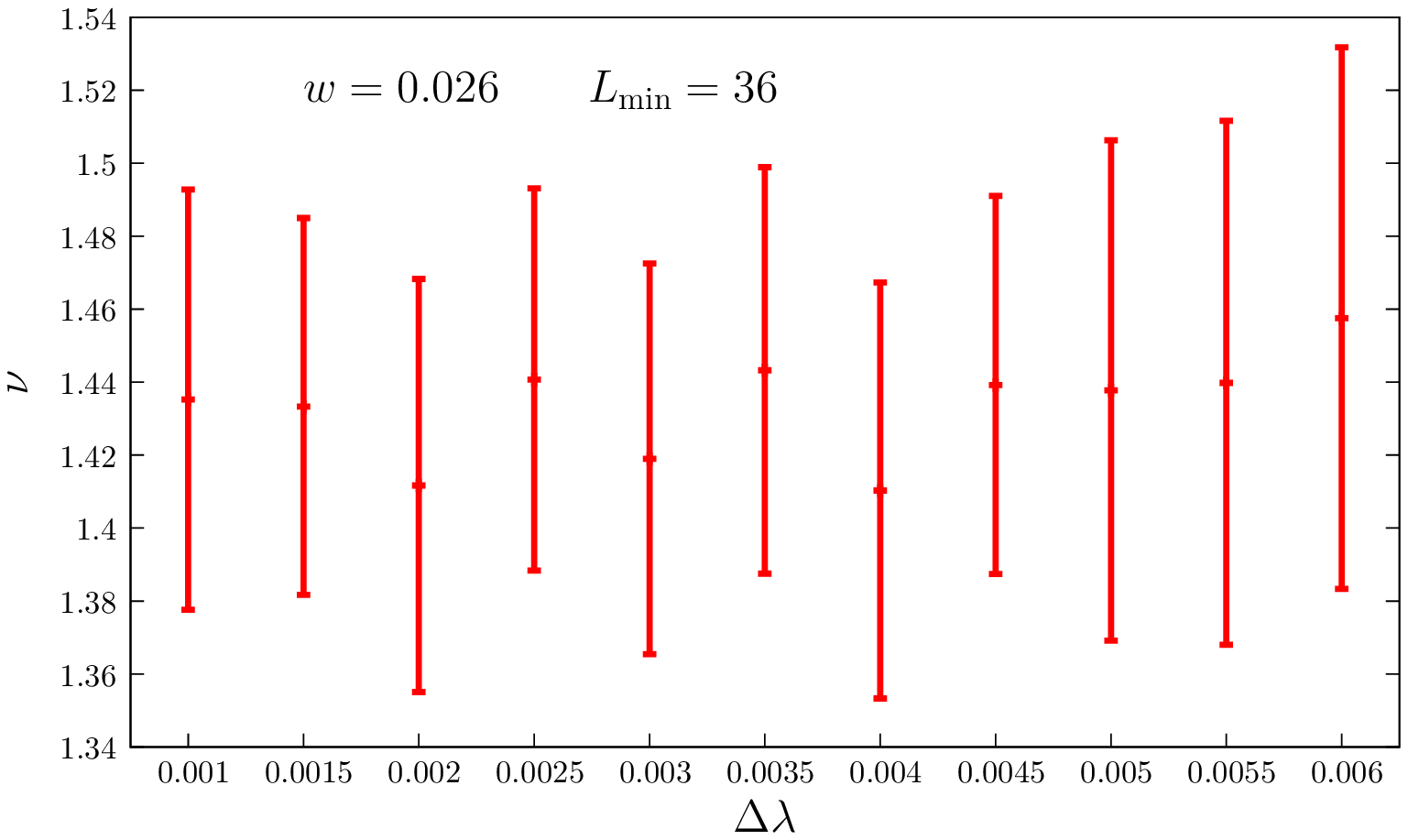}

\vspace{3ex}
\includegraphics[width=0.3\textwidth]{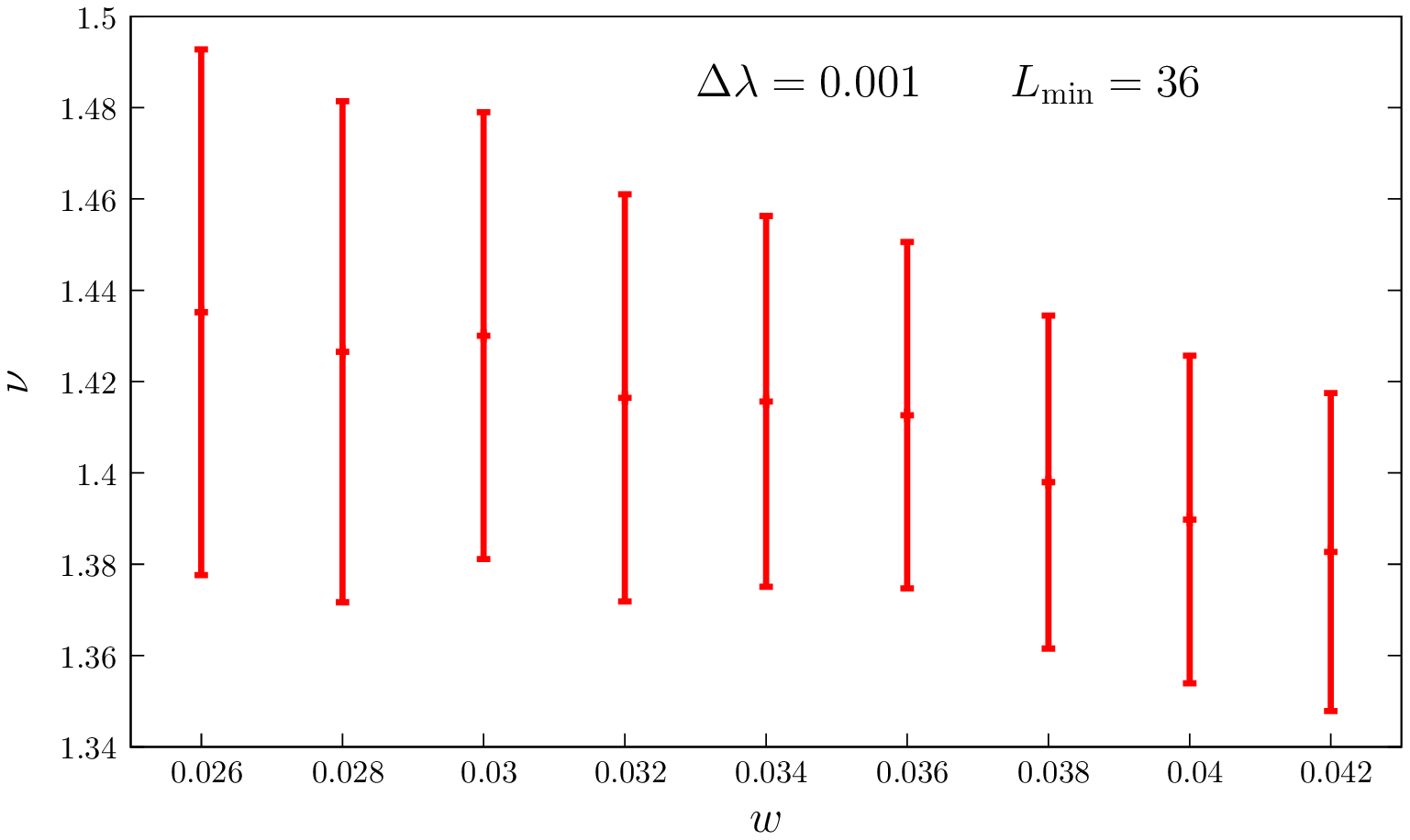}
  \caption{Dependence of the fitted value of $\nu$ on the bin size
    $\Delta\lambda$ for the smallest fitting
    range (upper panel) and on the width $w$ of
  the fitting range for the smallest bin size (lower panel). Here
  $L_{\rm min}=36$.} 
  \label{fig:binwidth_dep}
\end{figure}

In the Anderson model irrelevant operators are known to cause significant
finite size corrections to one-pa\-rameter scaling~\cite{corrections}. To see
how important that is in the present model, we performed the fits by omitting
the smallest volumes with system size $L<L_{\rm min}$. In
Fig.~\ref{fig:lmin_dep} we show the fitted value of $\nu$ as a function of the
smallest volume included in the fit (see also Table~\ref{tab:1}). Initially
the resulting value of $\nu$ increases with $L_{\rm min}$ but it stabilizes
around $L_{\rm min}=36$. A fit involving the leading irrelevant operator and
all the volumes, gives consistent results with this, indicating that finite
volume corrections are under control.  The fitting procedure we described
above also yields values for the critical point, $\lambda_c$. As a function of
$L_{\rm min}$, the fitted value of $\lambda_c$ shows no systematic dependence,
and different choices of $L_{\rm min}$ give consistent values within the
errors (see Table~\ref{tab:1}).

\begin{figure}[t!]
  \centering
  \includegraphics[width=0.42\textwidth]{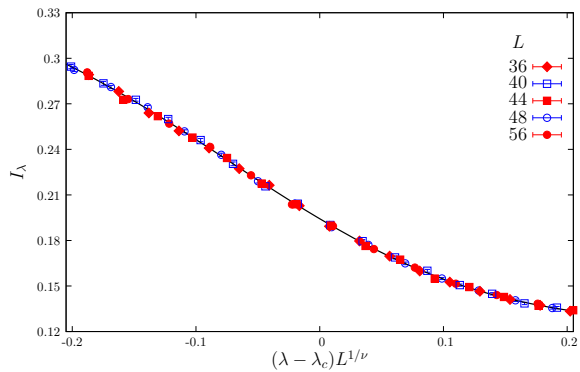}
  \caption{Integrated ULSD as a function of the scaling variable 
    $L^{1/\nu}(\lambda-\lambda_c)$ for several lattice sizes. The
    solid line is the approximate scaling function as obtained 
    through a fit to the data. Here $n_{\rm max}=9$ and $L_{\rm
      min}=36$, while $\Delta\lambda$ and $w$ were averaged over as
    explained in the text.}
  \label{fig:sf}
\end{figure}

There are two more arbitrary choices that can in principle affect the results.
These are the bin size, $\Delta\lambda$, over which the statistics for the ULSD
is collected and the width of the fitting range, $w$, around
$\lambda_c$.  We checked how these factors affect our results by
varying the bin size and the width of the fitting range, which we
always kept approximately centered at the critical point.  We
demonstrate both of these effects in Fig.~\ref{fig:binwidth_dep}. The
results show a slight tendency of $\nu$ to decrease as $\Delta\lambda$
is decreased, but it is rather stable for $\Delta\lambda\cdot
10^{3}\lesssim 3$. There is also a slight tendency of $\nu$ to
increase as $w$ is decreased, becoming rather stable for $w\cdot
10^{2}\lesssim 3$. To quote a single value for $\nu$, we averaged the
central values obtained for $1 \le \Delta\lambda\cdot 10^{3} \le 3$
and $2.6 \le w\cdot 10^{2} \le 3$. As the error is also rather stable
within these ranges, its average gives a good estimate of the typical
error, which we quote as the final error on $\nu$ for each choice of
$L_{\rm min}$. We have checked that other prescriptions (e.g.,
extrapolating to vanishing $w$ and/or $\Delta\lambda$, or changing
reasonably the ranges of $w$ and $\Delta\lambda$ over which the final
average is performed) give consistent results within the errors.

The fitting procedure described above also yields a polynomial parametrization
of the scaling function in the fitting range. To illustrate this, in
Fig.~\ref{fig:sf} we show the scaling function together with the data for the
range of system sizes used for the fit. Indeed, data from different volumes
collapse on a single scaling curve. 

Our final result for the critical exponent, $\nu=1.43(6)$, is compatible with
$\nu_{\rm U}=1.43(4)$ found earlier for the three-dimensional unitary Anderson
model~\cite{nu_unitary}. This strongly suggests that the transition in the
spectrum of the Dirac operator above $T_c$ is a true Anderson-type phase
transition, belonging to the universality class of the corresponding
three-dimensional Anderson model. Although its full physical implications are
not clear yet, localization might explain the large hadron screening masses
above $T_c$ and might have some implications for QCD-like theories that do
have a finite temperature chiral/deconfining phase transition.

%%%%%


\begin{thebibliography}{99}

%\cite{Anderson58}
\bibitem{Anderson58}
P.~W.~Anderson, Phys.\ Rev.\ {\bf 109}, 1492 (1958).

%\cite{Lee:1985zzc}
\bibitem{Lee:1985zzc}
  P.~A.~Lee and T.~V.~Ramakrishnan,
  %``Disordered electronic systems,''
  Rev.\ Mod.\ Phys.\  {\bf 57}, 287 (1985).
  %%CITATION = RMPHA,57,287;%%

%\cite{Evers:2008zz}
\bibitem{Evers:2008zz}
  F.~Evers and A.~D.~Mirlin,
  %``Anderson transitions,''
  Rev.\ Mod.\ Phys.\  {\bf 80}, 1355 (2008).
  %%CITATION = RMPHA,80,1355;%%

%\cite{KMOS}
\bibitem{KMOS}
B.~Kramer, A.~MacKinnon, T.~Ohtsuki and K.~Slevin, 
%``Finite size scaling analysis of the Anderson transition
Int.\ J.\ Mod.\ Phys.\ B {\bf 24}, 1841 (2010).

%\cite{Kovacs:2012zq}
\bibitem{Kovacs:2012zq} 
  T.~G.~Kov\'acs and F.~Pittler,
  %``Poisson to Random Matrix Transition in the QCD Dirac Spectrum,''
  Phys.\ Rev.\ D {\bf 86}, 114515 (2012).
%  [arXiv:1208.3475 [hep-lat]].
  %%CITATION = ARXIV:1208.3475;%%

%\cite{Verbaarschot:2000dy}
\bibitem{Verbaarschot:2000dy}
  J.~J.~M.~Verbaarschot and T.~Wettig,
  %``Random matrix theory and chiral symmetry in QCD,''
  Ann.\ Rev.\ Nucl.\ Part.\ Sci.\  {\bf 50}, 343 (2000).
%  [arXiv:hep-ph/0003017].
  %%CITATION = ARNUA,50,343;%%

%\cite{GarciaGarcia:2006gr}
\bibitem{GarciaGarcia:2006gr}
  A.~M.~Garc\'ia-Garc\'ia and J.~C.~Osborn,
  %``Chiral phase transition in lattice QCD as a metal-insulator transition,''
  Phys.\ Rev.\  D {\bf 75}, 034503 (2007).
%  [arXiv:hep-lat/0611019].
  %%CITATION = PHRVA,D75,034503;%%

%\cite{nu_latest}
\bibitem{nu_latest}
K.~Slevin and T.~Ohtsuki,  arXiv:1307.4483.

%\cite{offd_disorder} 
\bibitem{offd_disorder}
P.~Biswas, P.~Cain, R.~A.~R\"omer and M.~Schreiber, 
%``Off-Diagonal Disorder in the Anderson Model of Localization''
Phys.\ Status Solidi B, {\bf 218}, 205 (2000). 


%\cite{Bruckmann:2011cc}
\bibitem{Bruckmann:2011cc} 
  F.~Bruckmann, T.~G.~Kov\'acs and S.~Schierenberg,
  %``Anderson localization through Polyakov loops: lattice evidence and Random
  %matrix model,''
  Phys.\ Rev.\ D {\bf 84}, 034505 (2011).
%  [arXiv:1105.5336 [hep-lat]].
  %%CITATION = ARXIV:1105.5336;%%

%\cite{Montvay:1994cy}
\bibitem{Montvay:1994cy}
  I.~Montvay and G.~M\"unster,
  ``Quantum fields on a lattice,'' Cambridge, UK: Univ. Pr. (1994).

%\cite{Aoki:2005vt}
\bibitem{Aoki:2005vt} 
  Y.~Aoki, Z.~Fodor, S.~D.~Katz and K.~K.~Szab\'o,
  %``The Equation of state in lattice QCD: With physical quark masses towards
  %the continuum limit,''
  JHEP {\bf 0601}, 089 (2006);
%  [hep-lat/0510084];
  %%CITATION = HEP-LAT/0510084;%%
%
%\cite{Borsanyi:2010cj}
%\bibitem{Borsanyi:2010cj} 
  S.~Bors\'anyi, G.~Endr\H odi, Z.~Fodor, A.~Jakov\'ac, S.~D.~Katz, S.~Krieg,
  C.~Ratti and K.~K.~Szab\'o,
  %``The QCD equation of state with dynamical quarks,''
  JHEP {\bf 1011}, 077 (2010).
%  [arXiv:1007.2580 [hep-lat]].
  %%CITATION = ARXIV:1007.2580;%%

\bibitem{HS} E.~Hofstetter
    and M.~Schreiber, Phys. Rev. B {\bf 49},  14726 (1994).

\bibitem{SSSLS} 
    B.~I.~Shklovskii, B.~Shapiro,  
  B.~R.~Sears, P.~Lambrianides and H.~B.~Shore, Phys. Rev. B {\bf
    47}, 11487 (1993).

\bibitem{SP}  F.~Siringo and
    G.~Piccitto, J. Phys. A {\bf 31}, 5981 (1998).

%\cite{James:1975dr}
\bibitem{James:1975dr} 
  F.~James and M.~Roos,
  %``Minuit: A System for Function Minimization and Analysis of the Parameter
  %Errors and Correlations,''
  Comput.\ Phys.\ Commun.\  {\bf 10}, 343 (1975).
  %%CITATION = CPHCB,10,343;%%

%\cite{Lepage:2001ym}
\bibitem{Lepage:2001ym} G.~P.~Lepage, B.~Clark, C.~T.~H.~Davies,
  K.~Hornbostel, P.~B.~Mackenzie, C.~Morningstar and H.~Trottier,
  %``Constrained curve fitting,''
  Nucl.\ Phys.\ Proc.\ Suppl.\  {\bf 106}, 12 (2002).
%  [hep-lat/0110175].
  %%CITATION = HEP-LAT/0110175;%%

\bibitem{corrections}
K.~Slevin and T.~Ohtsuki, Phys.\ Rev.\ Lett.\ {\bf 82}, 382 (1999).

\bibitem{nu_unitary}
K.~Slevin and T.~Ohtsuki, Phys.\ Rev.\ Lett.\ {\bf 78}, 4083 (1997).

\bibitem{nu_symp} Y.~Asada, K.~Slevin and T.~Ohtsuki,
  {J. Phys. Soc. Jpn.} {\bf 74} supplement, 238 (2005).
%  [cond-mat/0410190 [cond-mat.dis-nn]].


\end{thebibliography}
\end{document}